\documentclass[prb,twocolumn,showpacs,preprintnumbers,amsmath,amssymb]{revtex4-1}
\usepackage{amssymb}
\usepackage{bm}
\usepackage{amsmath}
\usepackage[dvips]{graphicx}
\usepackage{color}
\usepackage{comment}
\usepackage{wrapfig}
\usepackage{textcomp}
\usepackage{multirow}

\begin{document}
\setlength{\fboxsep}{0pt}
\begin{abstract}
This work shows that a strongly correlated phase which is gapped to collective spin excitations but gapless to charge fluctuations emerges as a universal feature in one-dimensional fermionic systems obeying certain discrete symmetries whenever one pair of spin-degenerate subbands is occupied and an arbitrarily weak spin-orbit interaction is present. This general result is independent of the details of the one-dimensional confinement, the fermionic spin or nature of the spin-orbit interaction. In narrow-gap semiconductors, this gap may be of order 10 \textmu eV. This strongly correlated phase may be identified both via an anomalous $h/2e$ flux periodicity in Aharonov-Bohm oscillations and $2e$ periodic Coulomb blockade, features which reflect the existence of fermionic pairing despite the absence of superconductivity and the repulsive nature of the interaction.
\end{abstract}

	\title{Symmetry-protected spin gaps in quantum wires}
\author{Tommy Li}
\affiliation{Center for Quantum Devices, Niels Bohr Institute,
	University of Copenhagen, DK-2100 Copenhagen, Denmark}

\maketitle 

\section{Introduction}

Interacting fermions confined to one dimension (1D) are paradigmatic examples of strongly correlated systems, displaying  properties such as interaction-dependent critical exponents and spin-charge separation. The celebrated Luttinger liquid, which first appeared in models without backscattering, possesses a gapless spectrum of bosonic excitations \cite{Tomonaga,Luttinger,MattisLieb}. For spinful systems with backscattering, however, the ground state was shown by Luther and Emery  to depend on the sign of the interaction \cite{LutherEmery}. When spin rotational symmetry is present,  Luttinger liquid properties are maintained for repulsive interactions, while for attractive interactions, a new correlated state arises which is gapped to spin excitations but gapless to excitations of the total charge. Whereas Luttinger liquid properties are extremely challenging to identify in experiment, the predicted experimental signatures of the Luther-Emery phase are striking. The spin gap manifests as a vanishing of the single particle density of states at low energies \cite{VoitSpectral}, a flux periodicity of $2e$ indicative of fermionic pairing \cite{SeidelLee}, and vanishing backscattering from impurities at the Fermi level \cite{KainarisScattering}. These characteristics, which bear remarkable similarities with the superconducting state appear nevertheless in the absence of superconducting order.

A number of previous studies have demonstrated that the spin-gapped phase may be realized in the presence of purely repulsive electron-electron interactions in  certain multiband systems \cite{FinkelsteinLarkin,Cheng2011,Kraus2013,Lang2015,Guther2017,LBF} as well as Dirac semimental nanowires with an external magnetic field \cite{ZhangDS}. In this work, we will study an analogous situation in which the spin gap is induced by the spin-orbit interaction in two-band systems interacting via Coulomb repulsion, and emerges as a universal feature regardless of the specific model as long as long as certain discrete symmetries are obeyed.

\begin{figure}[t]
	\begin{tabular}{cc}
		\fbox{\includegraphics[width = 0.2\textwidth]{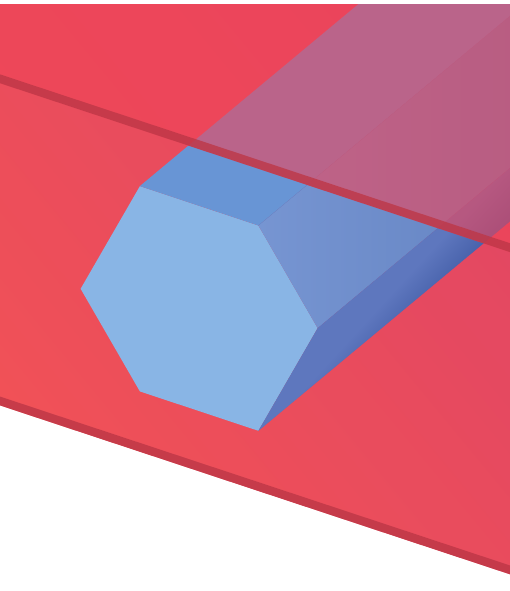}} &
		\fbox{\includegraphics[width = 0.2\textwidth]{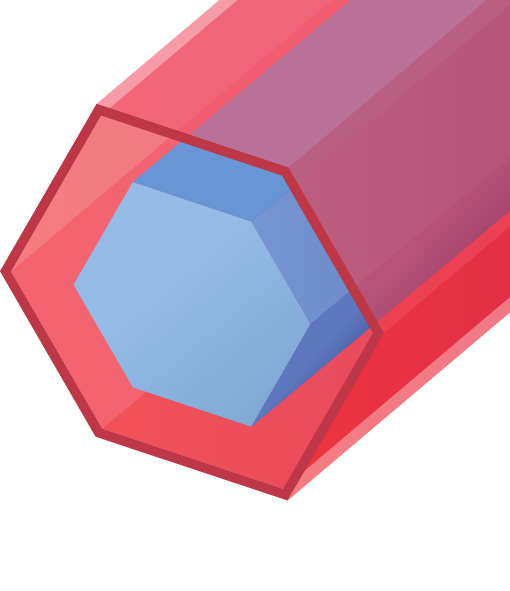}}  
	\end{tabular} 
	\caption{ Examples of possible geometries for screening of the nanowire: (left) two planes symmetrically placed around the wire, and (right) a single conductor enclosing the wire.}
	\label{fig:screening}
\end{figure}

\begin{figure}
	\includegraphics[width = 0.48\textwidth]{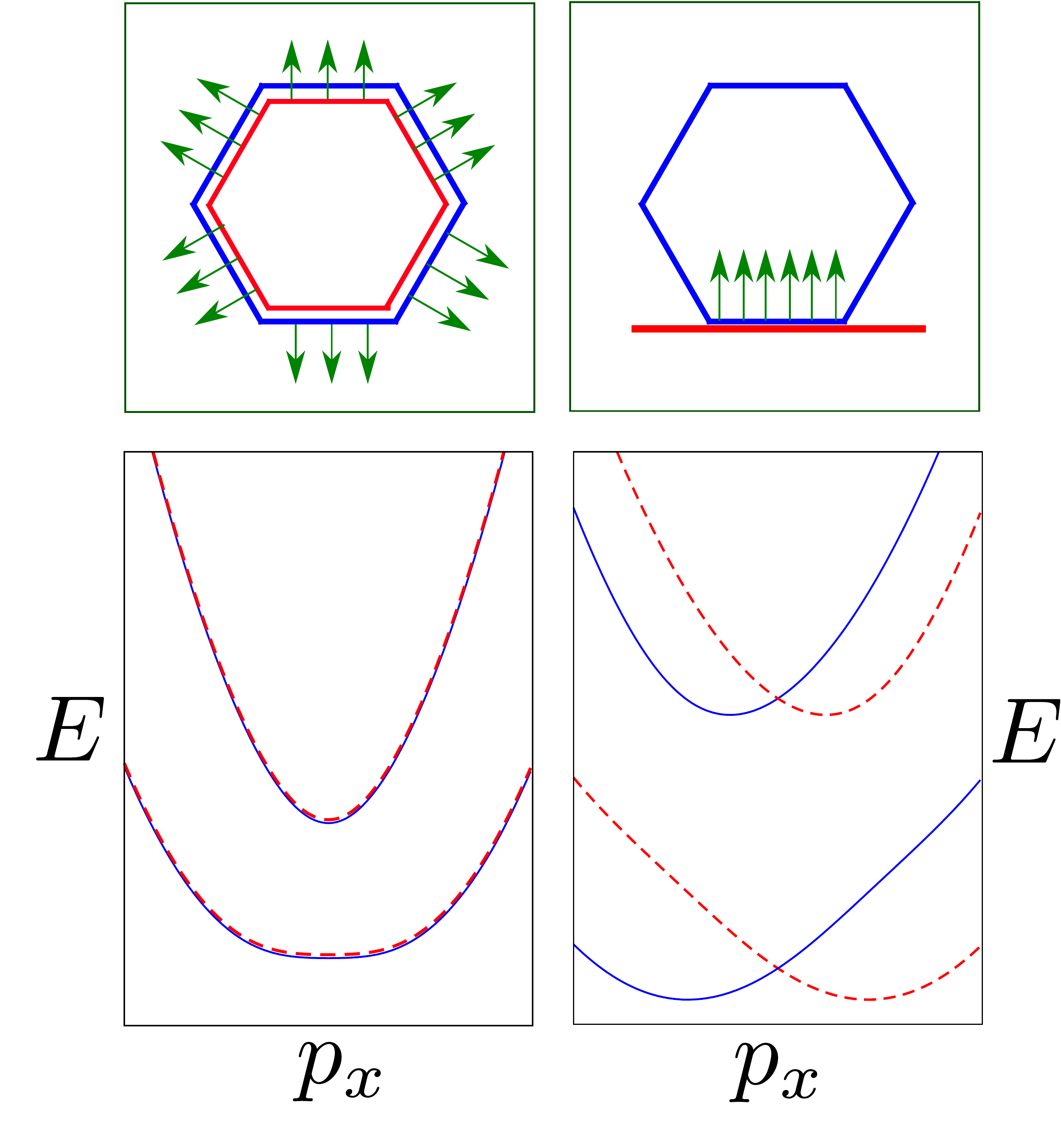}
	\caption{The single-electron dispersion for (left) inversion symmetric and (right)  asymmetric setups, with the top figures showing the cross-sectional view of the nanowire. In the left setup the nanowire rests directly on a substrate, while on the right it is fully surounded by a single material. The green arrows show the direction of the electric field (left) symmetrically radiating away from the core of the wire and (right) at the contact between a nanowire and a 2D plane.}
	\label{fig:disp}
\end{figure}

A concrete physical setup involving a hexagonal wire which serves as an example of the general theorem we will derive is shown in Fig. \ref{fig:screening}. The top two panels show free-standing hexagonal nanowires screened by either (left) two symmetrically placed conducting planes or (right) a single conductor enclosing the wire. Such systems exhibit spatial inversion symmetry by design. By contrast, previous theoretical studies on interacting nanowires with Rashba spin-orbit coupling have exclusively focused on the situation in which inversion symmetry is broken, focusing on Luttinger-liquid behaviour \cite{YuSC,MorozBarnesSC,ChengSC} or interaction effects associated with partial gapping of the single-particle dispersion in the presence of an external magnetic field \cite{MengLoss,SchmidtPedder,SunSDWSOC}, while one previous study showed that the Luther-Emery state could be realized above a critical interaction strength \cite{PedderLE}. In a common physical setup, one facet of the nanowire is contacted directly to a two-dimensional interface giving rise to a strong unidirectional electric field penetrating part of the surface. Alternatively, inversion asymmetry may arise if the wire is partially coated with a metal. These situations may be contrasted with an inversion symmetric setup in which the electric field at the surface of a free-standing core-shell nanowire points radially over the entire surface of the wire in a symmetric fashion. The difference between these physical setups is illustrated in the top panels of Fig. ~\ref{fig:disp}. The interaction effects differ significantly in the two cases due to the structure of the single-particle wavefunctions and dispersion, which are shown in the lower panels of Fig. ~\ref{fig:disp} for the lowest 1D subbands. In the symmetric case (shown left), the single particle dispersion possesses a twofold degeneracy, $\epsilon_\uparrow(k) = \epsilon_\downarrow(k)$ which is lifted when inversion symmetry is broken (shown right).

One specific model corresponding to the case shown in the left panels of Fig. ~\ref{fig:disp} is defined by the Hamiltonian (choosing coordinates so that $x$ points along the wire axis)
\begin{gather}
H_0(\bm{r},\bm{p},\bm{S}) = \frac{p_x^2 + p_\perp^2}{2m} + V(\bm{r}_\perp) + \underbrace{\lambda_R\nabla V \times \bm{p} \cdot \bm{S} }_{H_{SO}} \ \ ,
\label{H_so}
\end{gather}
where the first term represents the kinetic energy and the second term is the hexagonal 1D confining potential, with $\bm{r}_\perp$ being the coordinate vector in the plane perpendicular to the wire axis. The final term is the spin-orbit interaction, parametrized by the Rashba interaction strength $\lambda_R$, which generates a locally varying spin polarization (or spin texture) which winds around the cross section of the wire. The single-particle Hamiltonian is manifestly symmetric under the operator $\mathcal{P}$ of spatial inversion, $(x,y,z) \rightarrow (-x,-y,-z)$, $( p_x, p_y, p_z) \rightarrow (-p_x, -p_y, -p_z)$, $(S_x, S_y, S_z) \rightarrow(S_x, S_y, S_z)$. Due to the fact that time-reversal symmetry is also present, the single particle band structure is also twofold degenerate, $\epsilon_\uparrow(p_x) = \epsilon_{\downarrow}(p_x)$. This is in contrast to the situation shown in the right panels of Fig. \ref{fig:disp}, which violates spatial inversion symmetry since $V(\bm{r}_\perp) \neq V(-\bm{r}_\perp)$, leading to a lifting of the degeneracy, $\epsilon_\uparrow(p_x) \neq \epsilon_\downarrow(p_x)$.

In this work we consider a general situation in which three discrete symmetries are present: time-reversal, spatial inversion, and a two-fold combined spin and spatial rotation about the axis of the wire. Our Renormalization Group (RG) and bosonization analysis of the interacting model reveals that an arbitrarily weak spin-orbit interaction is sufficient to open a spin gap, regardless of the geometry of the wire or screening surfaces, the total spin of the system, or the nature of the spin-orbit interaction as long as only one pair of degenerate bands is occupied. The spin gap is a manifestation of the interplay between the spin texture of the wire and the Coulomb interaction, and exists in the presence of screening by symmetric conducting surfaces of arbitrary geometry. This result implies that the ground state of any clean semiconductor nanowire possessing intrinsic spin-orbit coupling exhibits a spin gap at low densities, as long as the engineering of the system does not introduce inversion asymmetry.

\section{General symmetry analysis}
Consider a system of fermions confined to 1D which we may generally describe by a long-wavelength effective Hamiltonian $H = \int{ \psi^\dagger(\bm{r}) H_0(\bm{r},\bm{p},\bm{S}) \psi(\bm{r}) d^3 \bm{r}} + H_{int}$ where the interaction term is
\begin{gather}
H_{int} = 
\frac{1}{2} \int{ U(\bm{r}',\bm{r}) \psi^\dagger(\bm{r}')\psi(\bm{r}')  \psi^\dagger(\bm{r}) \psi(\bm{r}) d^3 \bm{r} d^3 \bm{r}'} \ \ ,
\label{Hamil3D}
\end{gather}
where  $\bm{S}$ are the spin operators, the fermionic creation operators $\psi^\dagger(\bm{r})$ are $2|S| +1$-component spinors, and $U(\bm{r}',\bm{r})$ is the Coulomb interaction which is screened by conducting surfaces external to the wire. Our analysis encompasses both $n$-type and $p$-type  semiconductor quantum wires: in the former case, we take $|S| = \frac{1}{2}$, while for the latter, due to the fact that the valence band states are formed from $p_{\frac{3}{2}}$ orbitals, the spin-orbit interaction must be described by a four-component spinor ($|S| = \frac{3}{2}$) which combines both the spin and atomic angular momentum degrees of freedom. We also do not assume that the spin-orbit interaction is linear in momentum (as in Eq. \ref{H_so}); in the case of $p$-type semiconductor wires the Hamiltonian may also contain terms which are quadratic in the spin operators \cite{LuttingerKohn}. In the absence of external magnetic fields, we may assume that $H_0$ is symmetric under time reversal ($\mathcal{T}$). We shall study the case where the Hamiltonian is symmetric under both spatial inversion $\mathcal{P}$ and combined spatial and spin rotation by an angle $\pi$ about the wire axis. We may then classify degenerate time-reversed pairs of single-particle states $|k, \uparrow\rangle, ~ |-k, \downarrow\rangle$ via the momentum $k$ along the nanowire axis and the eigenvalues $e^{i S \pi},  e^{-i S \pi}$ under the combined spin and spatial $\pi$-rotation.  We may express the single particle wavefunctions in a basis of spin states $|s\rangle$ with polarization along the wire axis (with $|k,\downarrow \rangle = \mathcal{P} \mathcal{T}| k,\uparrow \rangle$),
\begin{align}
& \langle \bm{r},  s |k,\uparrow \rangle  = e^{ikx}  f_{ks} ( \bm{r}_\perp) , \nonumber \\
& \langle \bm{r}, s|k,\downarrow \rangle =  e^{ikx}  f^*_{k,- s}(\bm{r}_\perp)  \ ,
\label{chi}
\end{align}
and  $f_{ks}(\bm{r}_\perp) = f_{ks}(y,z)$ are functions of the coordinates in the cross section, which satisfy
\begin{align}
f_{ks} (-\bm{r}_\perp) = f_{-k,s}(\bm{r}_\perp) =  (-1)^{S - s}  f_{ks}(\bm{r}_\perp) \ \ .
\end{align}

In the following analysis the interaction term in the Hamiltonian is required to satisfy both translational invariance along the wire axis and $\pi$-rotational symmetry in the cross section. Since the Coulomb interaction is spatially isotropic these symmetries can only be violated by the geometry of screening planes near the wire. Two examples of symmetric arrangements are shown in the  Fig. ~\ref{fig:screening}. Alternatively, the system may be screened only by a remote conducting surface at distance $L$ larger than both the dimensions of the cross section and the Fermi wavelength $2\pi/k_F$. In all cases the conducting surfaces are required to be homogeneous along the wire direction. For a remote screening plane parallel to the wire, the Coulomb interaction consists simply of the unscreened interaction $\propto 1/|\bm{r} - \bm{r}'|$ in addition to the potential due to an image charge at distance $2L$ from the wire and is axially isotropic. In the remaining cases the Coulomb interaction may be expanded in a basis of solutions $\varphi_{n,q}(\bm{r}) = \varphi_n(x,y) e^{iqx}$ of the Helmholtz equation which vanish on the screening surfaces, and the interaction  is given by
\begin{gather}
U(x', \bm{r}'_\perp, x,\bm{r}_\perp) = \int{ \sum_n{ \frac{ \varphi^*_n(\bm{r}_\perp') \varphi_n(\bm{r}_\perp)}{ \kappa_n^2 + q^2}} e^{iq(x- x')} \frac{dq}{2\pi}}\ \ ,
\label{int1}
\end{gather}
where $\kappa_n^2 +q^2$ are the eigenvalues of the Laplace operator. We will choose a basis in which $\varphi_n(\bm{r})$ are real. The first Born amplitudes $\langle f | H_{int} | i \rangle$ for scattering between initial and final two particle states $|i\rangle = |k_1 \alpha_1\rangle \otimes | k_2 \alpha_2 \rangle$, $|f\rangle = | k_3 \alpha_3 \rangle \otimes | k_4 \alpha_4 \rangle$ with initial and final momenta at the Fermi points, $|k_1| =|k_2| =|k_3| = |k_4| = k_F$ and $\alpha_1, \alpha_2 , \alpha_3, \alpha_4 = \uparrow, \downarrow$ may be related to the harmonics $\varphi_n$ and the overlaps of the spinor states $\chi_{k\alpha} (\bm{r})= \sum_s{ f_{ks}(\bm{r}) | s \rangle}$ via
\begin{align}
\langle f | H_{int} | &i \rangle = U_{k_3 \alpha_3, k_4 \alpha_4, k_1 \alpha_1, k_2 \alpha_2 } \nonumber \\
=& \sum_n{
	\frac{ F_{k_3 \alpha_3, k_1 \alpha_1}(n) F_{k_4 \alpha_4, k_2 \alpha_2}(n)}{ |k_3 - k_1|^2 + \kappa_n^2} 
} \ \ , \nonumber \\
F_{k \alpha, k' \beta}&(n) = \int{ \varphi_n(\bm{r}) \chi^\dagger_{k \alpha}(\bm{r}) \chi_{k' \beta}(\bm{r}) d^2 \bm{r}} \ \ ,
\label{int2}
\end{align}
where $\bm{r} = (y,z)$ and the inner products are given in terms of the spin components (\ref{chi}) via (writing $\chi_{+k_F,\alpha} \rightarrow \chi_{R, \alpha}$, $\chi_{-k_F, \alpha} \rightarrow \chi_{ L,\alpha}$)
\begin{align}
& \chi^\dagger_{R \alpha} \chi_{R \alpha} = \chi^\dagger_{ L \alpha} \chi_{ L \alpha} = \sum_s{ | f_{k_F,s} |^2} \ \ , \nonumber \\
& \chi^\dagger_{L\alpha} \chi_{R \alpha} = \sum_s{ (-1)^{ s - S} | f_{k_F,s}|^2} \ \ , \nonumber \\
& \chi^\dagger_{R\downarrow} \chi_{R\uparrow} = - \chi^\dagger_{L \downarrow} \chi_{L\uparrow} =  \sum_ s{ f_{k_F,s}  f_{k_F, -s} } \ \ , \nonumber \\
& \chi^\dagger_{ L\downarrow} \chi_{R \uparrow} = 0 \ \ .
\label{overlap}
\end{align}

The low-energy physics of the interacting system is determined by interaction processes involving electrons close to the Fermi level. We will analyze the situation when only the lowest degenerate pair of bands is occupied, so the only interactions which may become relevant in the infrared limit involve right- and left- moving particles in the $\uparrow, \downarrow$ spin bands. We introduce an effective field-theoretical  description of the model via chiral one-dimensional fermionic fields associated with each spin species $\psi_{R,\alpha}(x), \psi_{L,\alpha} (x)$ and express the interaction Hamiltonian as $H_{int} = \int{ \mathcal{H}_{int}(x) dx}$ which consists of all possible scattering processes consistent with the discrete symmetries and takes the form
\begin{gather}
\mathcal{H}_{int} = \frac{U_{R\uparrow,R\uparrow,R\uparrow,R\uparrow}}{2} \sum_{\alpha}{\left[ n_{R,\alpha}^2 + n_{L,\alpha}^2 \right]} \nonumber \\
+ (U_{R\uparrow,R\downarrow,R\uparrow,R\downarrow} - U_{R\uparrow,R\downarrow,R\downarrow,R\uparrow}) \left[ n_{R\uparrow} n_{R\downarrow} + n_{L\uparrow} n_{L\downarrow} \right] \nonumber \\
+ ( U_{R\uparrow,L\uparrow,R\uparrow,L\uparrow} - U_{R\uparrow,L\uparrow,L\uparrow,R\uparrow}) \left[ n_{R\uparrow} n_{L\uparrow} + n_{R \downarrow} n_{L\downarrow} \right] \nonumber \\
+ U_{R\uparrow,L\downarrow,R\uparrow,L\downarrow} \left[ n_{R\uparrow}n_{L\downarrow} + n_{R\downarrow} n_{L\uparrow} \right] \nonumber \\
+ (U_{R\uparrow,L\downarrow,R\downarrow,L\uparrow} - U_{R\uparrow,L\downarrow,L\uparrow,R\downarrow}) \sum_\alpha{
\psi^\dagger_{R,\alpha} \psi^\dagger_{L,-\alpha} \psi_{L,\alpha} \psi_{R,-\alpha} 
} \nonumber \\
+ U_{R\uparrow,L\uparrow,R\downarrow,L\downarrow} \sum_\alpha{
\psi^\dagger_{R,\alpha} \psi^\dagger_{L,\alpha} \psi_{L,-\alpha} \psi_{R,-\alpha} 
}
\label{fermionicfield}
\end{gather}
with $n_{R\alpha} = \psi^\dagger_{R,\alpha} \psi_{R,\alpha}$, $n_{L\alpha} = \psi^\dagger_{L,\alpha} \psi_{L,\alpha}$.

In the bosonic mapping  \cite{Giamarchi2003} we introduce fields $\theta_i, \phi_i$ associated with long-wavelength fluctuations of the total and relative band densities $n_{\uparrow} + n_\downarrow$, $n_\uparrow - n_\downarrow$ which we refer to as charge and spin densities respectively. We obtain the bosonized Hamiltonian from (\ref{fermionicfield}) via the operator substitutions $\psi_{R,\alpha} = \sqrt{\frac{\Lambda}{v_F}} e^{\frac{i}{2} ( \phi_\rho + \theta_\rho + \alpha  \phi_\sigma + \alpha \theta_\sigma)}$, $\psi_{L,\alpha} = \sqrt{ \frac{ \Lambda}{v_F}} e^{\frac{i}{2} ( \phi_\rho - \theta_\rho + \alpha \phi_\sigma - \alpha \theta_\sigma)}$ with $\alpha = +1, -1$ corresponding to $\uparrow,\downarrow$ respectively, and $\Lambda$ is a UV cutoff which we may consider to be the running RG energy scale. The spin modes are described by a Sine-Gordon theory while the charge modes are massless and the Hamiltonian density of the bosonized theory has the form
\begin{align}
\mathcal{H} = &\frac{v_\rho}{8\pi K_\rho} ( \partial_x \phi_\rho)^2 + \frac{ K_\rho v_\rho}{8\pi} ( \partial_x \theta_\rho)^2 \nonumber \\
+&\frac{v_\sigma}{8 \pi K_\sigma}( \partial_x \phi_\sigma)^2 + \frac{K_\sigma v_\sigma}{8\pi} ( \partial_x \theta_\sigma)^2 \nonumber \\
 +&2g \cos \sqrt{2} \phi_\sigma + 2 g'\cos \sqrt{2} \theta_\sigma \ 
\label{SG} \ \ ,
\end{align}
with the parameters $v_\rho, v_\sigma, K_\rho, K_\sigma$ related to the coefficients of the interaction terms in (\ref{fermionicfield}) involving the density operators $n_{R\alpha}, n_{L\alpha}$ while $g$ and $g'$ are proportional to the prefactors of the final two terms in (\ref{fermionicfield}).

The spin and charge sectors are  decoupled as a result of the combined time reversal and inversion symmetry. While $\mathcal{H}_\rho$ is harmonic, $\mathcal{H}_\sigma$ contains competing interactions $\cos \sqrt{2} \phi_\sigma$, $\cos \sqrt{2} \theta_\sigma$ originating from processes  in (\ref{fermionicfield}) of the form $\psi^\dagger_\uparrow \psi^\dagger_\downarrow \psi_\uparrow \psi_\downarrow$ and $\psi^\dagger_\uparrow \psi^\dagger_\uparrow \psi_\downarrow \psi_\downarrow$ respectively, which one might expect to generate several possible interacting phases depending on the values of the parameters $K_\sigma, g, g'$.  However, the symmetries of the three-dimensional spin texture associated with the spinor wavefunctions $\chi_{k\alpha}(y,z) = \sum_s{ f_{ks}(y,z) | s\rangle}$ in combination with the repulsive nature of the Coulomb interaction will enforce strict relations between the values of these parameters, resulting in universal low-energy properties. This situation persists in the presence of screening by symmetrically arranged conductors (Fig. ~\ref{fig:screening}). Excitations of $\phi_\sigma$ will always be gapped and the ground state contains a nonvanishing expectation value $\langle \cos \sqrt{2} \phi_\sigma \rangle \neq 0$.
 
The proof of this theorem proceeds straightforwardly from the relations between dimensionless interaction parameters of the bosonic theory and the  first-order fermionic scattering amplitudes. We have
\begin{gather}
K_\sigma =  \sqrt{\frac{1 + \hat{\Gamma}_\sigma - \Gamma_\sigma}{1 + \hat{\Gamma}_\sigma + \Gamma_\sigma} } \ \  ,\nonumber \\
\hat{\Gamma}_\sigma = \frac{U_{R\uparrow, R \downarrow, R\downarrow, R\uparrow}}{2\pi v_F} \ , \ 
\Gamma_\sigma = - \frac{U_{R \uparrow, L \uparrow, L \uparrow, R \uparrow}}{2\pi v_F} \ \ ,
\end{gather}
and (\ref{int2},\ref{overlap}) imply $\Gamma_\sigma < 0$ and $K_\sigma > 1$. The interaction proportional to $g'$ arises from scattering processes of the form $\psi^\dagger_\uparrow \psi^\dagger_\uparrow \psi_\downarrow \psi_\downarrow + h.c.$ and is given by
\begin{gather}
g' = \frac{U_{R \uparrow, L\uparrow, R\downarrow, L\downarrow}}{2\pi v_F} = \frac{1}{2\pi v_F}\sum_n{ \frac{F_{R\uparrow,R\downarrow}(n) F_{L\uparrow,L\downarrow}(n)}{ \kappa_n^2}}
\label{dualcondition}
\end{gather}
while the interaction $g$ arises from the sum of interfering forward and backward scattering processes $ g = \frac{  U_{R\uparrow,L\downarrow,R\downarrow,L\uparrow} - U_{R\uparrow,L\downarrow,L\uparrow,R \downarrow}}{2\pi v_F} =\frac{ U_{R\uparrow,L\downarrow,R\downarrow,L\uparrow}}{2\pi v_F} +  \Gamma_\sigma$, so
\begin{gather}
g - \Gamma_\sigma =  \frac{1}{2\pi v_F}
\sum_n{\frac{ F^*_{R\downarrow,R\uparrow}(n) F_{L\downarrow,L\uparrow}(n)}{\kappa_n^2}} < 0 \ \ , 
\label{LEcondition}
\end{gather}
and the inequality follows from the fact that the local overlap of spinor states $\chi^\dagger_{L\downarrow}(\bm{r})\chi_{L\uparrow}(\bm{r}) = - \chi^\dagger_{R\downarrow}(\bm{r})\chi_{R\uparrow}(\bm{r})$ is odd under both inversions of the spatial coordinate $\bm{r} \rightarrow -\bm{r}$ and longitudinal momentum, $L \leftrightarrow R$ (\ref{overlap}), a result of the winding of the spin texture in the cross section of the wire. The summation in  (\ref{LEcondition}) is performed over terms which are real and negative, while in general the summand in the expression for $g'$ (\ref{dualcondition}) is complex, so that we have strictly $|g'| \leq |g - \Gamma_\sigma|$.

\begin{figure}
	\includegraphics[width = 0.45\textwidth]{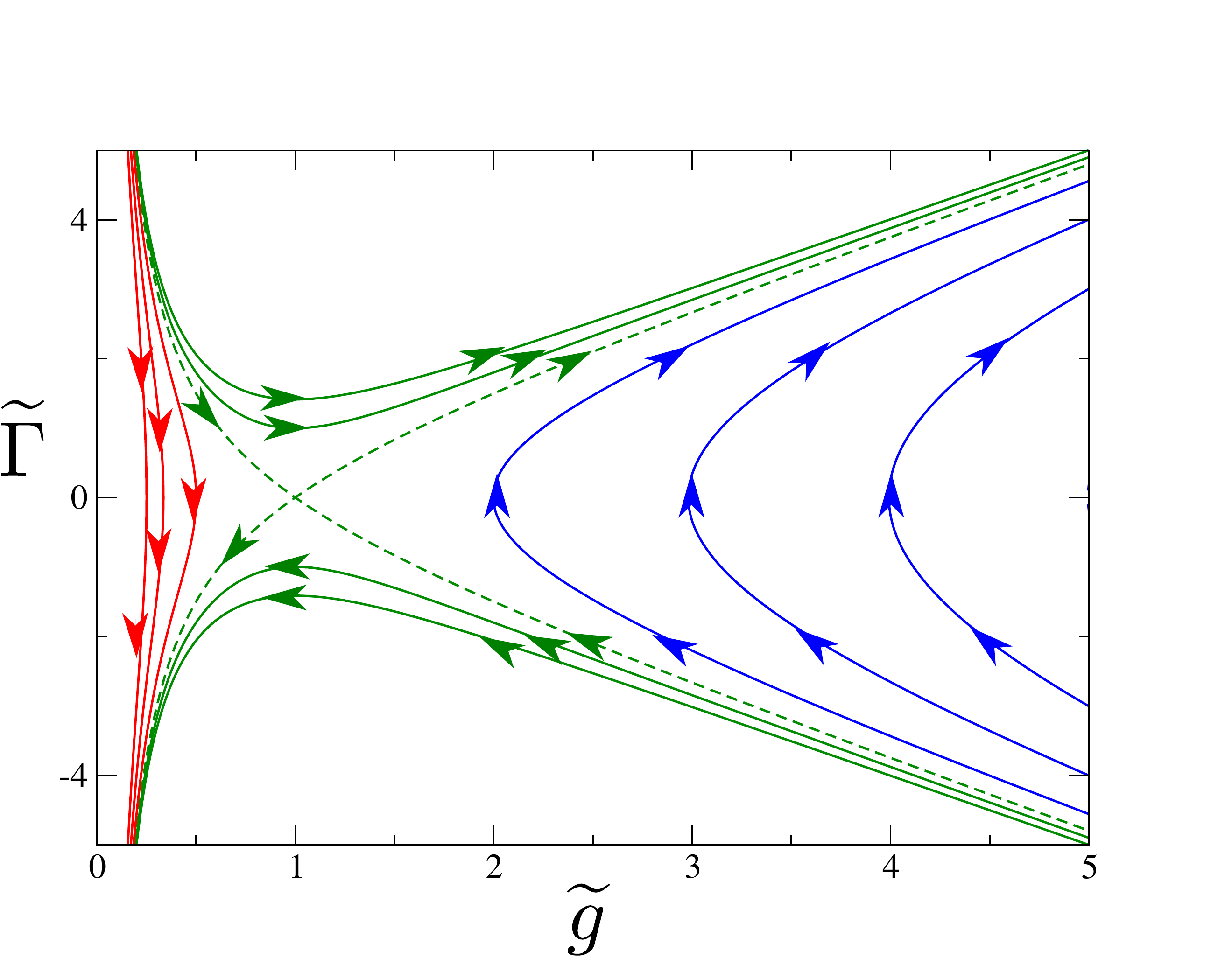}
	\caption{The RG flow of the couplings $\widetilde{\Gamma} = \Gamma_\sigma/\sqrt{|g_0 g_0'|}$, $\widetilde{g} = |g|/\sqrt{|g_0 g_0'|}$.}
	\label{fig:RG}
\end{figure} 

The scaling relations for the couplings to second order are given by
\begin{align}
&\frac{ dg}{dl} = 2 \Gamma_\sigma g , \  \ \ 
\frac{d g'}{dl} = - 2 \Gamma_\sigma g' , \nonumber \\
&\frac{d \Gamma_\sigma}{dl} = 2(g^2 - (g')^2)  ,
\label{RG}
\end{align}
with $l = \ln (\Lambda/\mu)$ where $\Lambda \approx E_F$ is the UV cutoff and $\mu$ is an energy scale which runs toward the infrared limit. If either the interactions $g,g'$ are initially zero, they remain zero under the RG flow. This situation occurs, for example, when the nanowire possesses an additional rotational symmetry whose selection rules forbid interactions of the form $\psi^\dagger_\uparrow \psi^\dagger_\uparrow \psi_\downarrow \psi_\downarrow$ and therefore $g' = 0$. In this case the RG equations may be integrated easily, and the system is gapped when the bare couplings satisfy $|g| > |\Gamma_\sigma|$, a condition which is always fulfilled due to the sign of the matrix element (\ref{LEcondition}). Solution of (\ref{RG}) then shows that the system flows to strong coupling at an energy scale $l = l^*$, which indicates the presence of a spin gap given by
\begin{gather}
\Delta = \Lambda e^{- l^*} \ , \nonumber \\
l^* = \frac{ 1}{ 2\sqrt{g^2 - \Gamma_\sigma^2}} \left[ \frac{\pi}{2} + \tan^{-1} \frac{ |\Gamma_\sigma|}{ \sqrt{g^2 - \Gamma_\sigma^2}} \right] 
\label{gap}
\end{gather}
where $\Gamma_\sigma, g$ represent the bare value of the couplings.

\begin{figure}
	\includegraphics[width = 0.5 \textwidth]{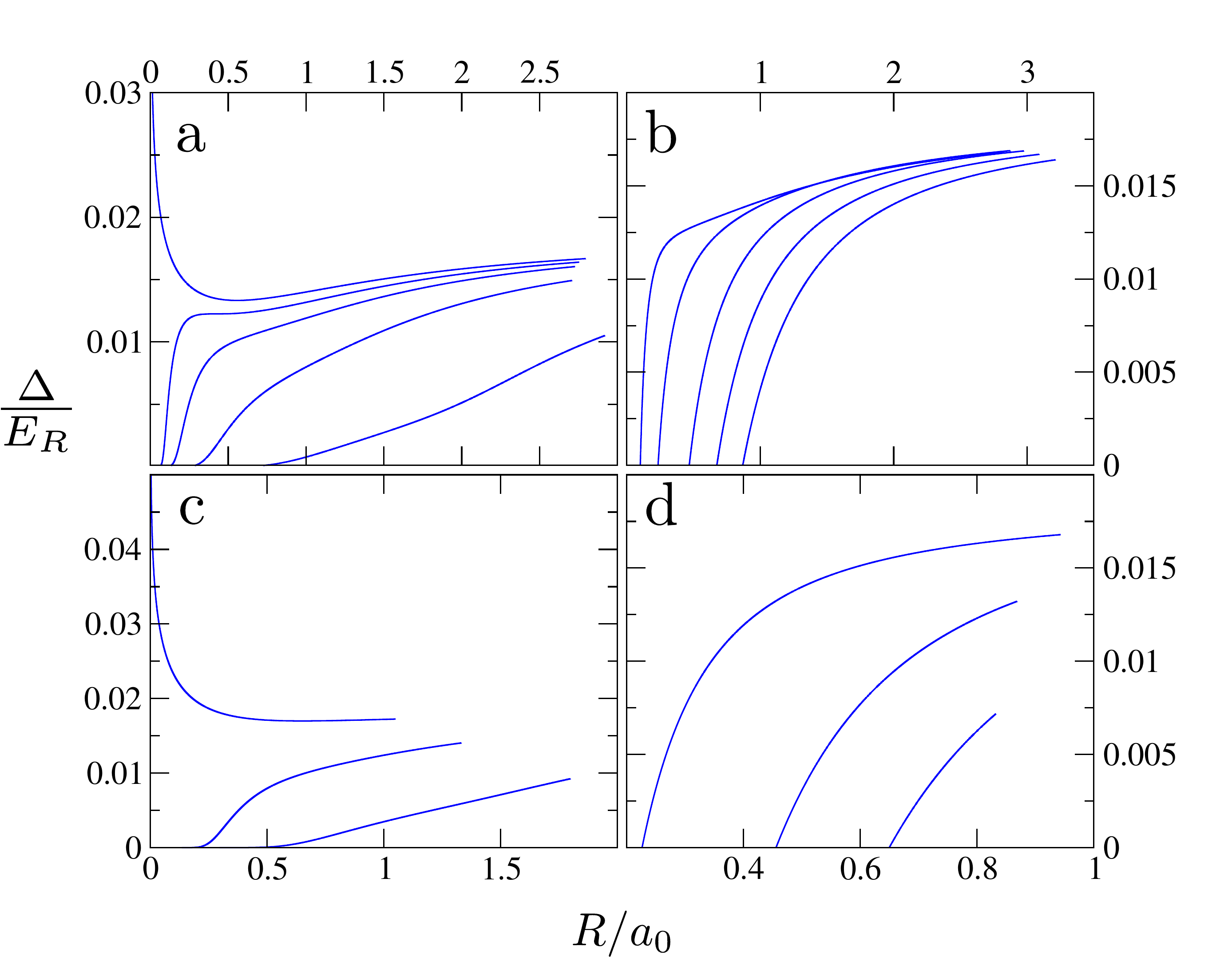}
	\caption{Calculated values of $\Delta$ (\ref{gap}) at values of the density at the fixed ratio $\Delta/E_F = e^{-3} \approx 1/20$, for various values of $b$, and plotted as a function of the wire radius $R/a_0$. The values are:  (a) $b = 0.8, 1.2, 1.4, 1.5, \frac{1+ \sqrt{5}}{2}\approx 1.618$, uniformly increasing with $b$, (b) $b = 1.7, 1.8, 2, 2.2, 2.4$, uniformly decreasing with $b$, (c) $b = -0.4, -0.5, -0.618$, uniformly decreasing with $b$ (bold), (d) $b = -0.7, -0.8, -0.9$, uniformly increasing with $b$. At special values $b = \frac{1 \pm \sqrt{5}}{2}$, the gap diverges in the limit of small wire radius. The curves terminate at values of the wire radius for which the  RG scale $l^* < 3$ for all values of the chemical potential below the second pair of degenerate subbands. These results are independent of material parameters.}
	\label{fig:delta}
\end{figure}

For the case when $g_0' \neq 0$, the RG equations exhibit a duality relation $gg' = const. = g_0 g_0'$, so that the flow to strong coupling of one of either couplings implies the vanishing of the other. The RG flows of the couplings $\widetilde{g} = |g|/\sqrt{|g_0 g_0'|}$,~~ $\widetilde{\Gamma} = \Gamma_\sigma/\sqrt{|g_0 g_0'|}$ are plotted in Fig. ~\ref{fig:RG}.  The phase diagram consists of four regions separated by the dashed lines, with a fixed point at $(\widetilde{g},\widetilde{\Gamma} )= (1,0)$. Comparison of Eqs. (\ref{dualcondition},\ref{LEcondition}) implies that $|g'|< |g - \Gamma_\sigma| \rightarrow |\widetilde{\Gamma}| < \widetilde{g} - 1/\widetilde{g}'$, thus the couplings $(\widetilde{g},\widetilde{\Gamma})$ initially lie above the lower separatrix (dashed line in Fig. \ref{fig:RG}), and flow in the portion of the phase diagram indicated by the blue lines towards strong coupling, $\Gamma_\sigma \rightarrow +\infty$, $|g| \rightarrow + \infty$, $g' \rightarrow 0$. We therefore find that the spin sector is gapped for both $g' = 0$ and $g' \neq 0$ and the ground state develops a nonvanishing expectation value $\langle \cos \sqrt{2} \phi_\sigma \rangle$. While this gap may occur generally in models with appropriately tuned values of the couplings, in our situation the gap is protected by the symmetry of the wavefunctions (\ref{overlap}) which involves both spin and orbital degrees of freedom. The gap ultimately arises from the negative sign of the matrix element (\ref{LEcondition}) which provides an effective attraction despite the underlying Coulomb interaction being repulsive.

\section{Nanowires with Rashba interaction}

The most promising experimental candidates for observation of the spin gap are free standing hexagonal nanowires in nanowires in narrow gap superconductors (e.g. InAs or InSb), which possess a strong intrinsic spin-orbit interaction. For $n$-type doping the system is described by the Hamiltonian (\ref{H_so}). For a hexagonal core-shell wire we may approximate the confining potential $V(\bm{r}_\perp) = \phi(|\bm{r}_\perp|) = \phi(r_\perp)$ with a circularly symmetric quantum well  which strongly localizes the charge density at the surface of the wire. The spin-orbit interaction $H_{SO} \propto \phi'(r_\perp) \bm{L} \cdot \bm{S}$ conserves the sum of orbital angular momentum and spin, and thus the eigenstates of the Hamiltonian possess definite $\bm{L} + \bm{S}$ along the direction of the wire. It follows that, by symmetry considerations, the states $|k,\alpha \rangle$ are of the form
\begin{align}
&\langle \bm{r} | k,\uparrow\rangle = \varphi(r_\perp) e^{i kx} \left[ 
\cos \frac{\xi_k}{2} | \frac{1}{2} \rangle -i \sin \frac{\xi_k}{2}  e^{i \theta} | -\frac{1}{2} \rangle 
 \right]  \nonumber \\
& \langle \bm{r} | k,\downarrow \rangle = \varphi(r_\perp) e^{ikx} \left[
\cos \frac{\xi_k}{2} | -\frac{1}{2} \rangle +i \sin \frac{\xi_k}{2} e^{-i \theta} | \frac{1}{2} \rangle 
\right]
\label{rashbawf}
\end{align}
where $\theta$ the angular coordinate in the cross-sectional plane and $\varphi(r_\perp)$ is a function which vanishes away from the surface of the wire. Acting on the wavefunction (\ref{rashbawf}) with the Hamiltonian (\ref{H_so}) we find that
\begin{gather}
\xi_k = \frac{2k R \beta}{ \frac{1}{2m R^2} + \beta}
\end{gather}
where
\begin{gather}
\beta = -\frac{ \lambda_R}{2R^2} \int{ \varphi(r)^2 \phi'(r) r dr } \ \ ,
\end{gather}
and the dispersion is given by
\begin{gather}
\epsilon(k) = \frac{k^2}{2m} 
+\frac{1}{2} \sqrt{(2\beta k R)^2 + ( \frac{1}{2m R^2} + \beta)^2} \ \ .
\end{gather}

For simplicity we may consider the case where the nearest screening plane is sufficiently far from the wire that the  interaction is isotropic and effectively unscreened over lengths comparable to the nanowire radius. Then due to rotational symmetry,  $g' = 0$. The gap (\ref{gap}) depends on the dimensionless parameter $b = 2m R^2 \beta$, the Fermi energy as well as the strength of the interaction, which we may parametrize via the effective Bohr radius $a_0 = \epsilon_r/me^2$ and Rydberg energy $E_R = me^4/2\epsilon_r^2$. The perturbative expression (\ref{gap}) is quantitatively accurate for $\Delta \ll \Lambda \approx E_F$. As the UV scale $\Lambda$ is increased, the density of states is reduced, diminishing the exponential factor; (\ref{gap}) rougly predicts a maximum value $\Delta < E_F e^{-2}$. Fig. ~\ref{fig:delta} shows the value of the gap when  $E_F$ is tuned so that $l^* = 3$ at fixed  radius $R/a_0$ and lies below the edge of the upper 1D subbands (thus $\Lambda$ and $\Delta$ are separated by a factor of $e^3 \approx 20$ and the perturbative RG analysis is valid). At fixed $R$, the gap is maximum for values $b = \frac{1 \pm \sqrt{2}}{2}$  and diverges in the limit of small $R$. In InAs, the parameters $a_0 = 35~\text{nm},~~E_R = 1.4\text{meV}$ and in InSb, $a_0 = 63~\text{nm},~~E_R = 0.68\text{meV}$, thus the typical nanowire radius $R \approx 50~\text{nm}$ is comparable to $a_0$. We then obtain  maximum gaps $\Delta \approx E_F/(e^3) \approx $ 20\textmu eV and 24\textmu eV for values $b = \frac{1 + \sqrt{5}}{2}, \frac{1- \sqrt{5}}{2}$ respectively for InAs, and $\Delta \approx $ 9, 12\textmu eV in InSb.

\begin{figure}
	\includegraphics[width = 0.3\textwidth]{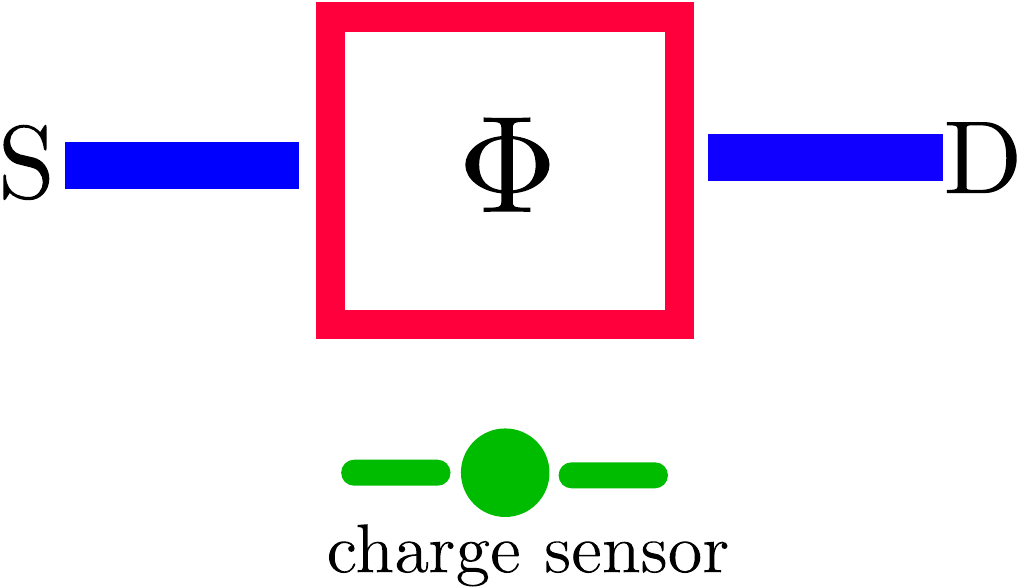}
	\caption{A loop geometry with tunnelling contacts (S,D) for  detection of the spin gap. Four nanowires are arranged in a planar loop (red). Aharanov-Bohm oscillations of the conductance will exhibit a $h/2e$ periodicity with respect to the flux $\Phi$ threading the loop. A charge sensor may be used to measure Coulomb blockade through the device as the density in the loop is tuned by symmetrically placed gates (not shown) some distance above and below the plane of the loop.}
	\label{fig:loop}
\end{figure}

\section{Experimental detection}
The existence of a spin gap may be detected via tunnelling experiments in a four-nanowire loop setup shown in Fig. \ref{fig:loop}. Since tunnelling of single electrons from the leads will change the spin of the system, this process is forbidden when the bias is below the gap. To lowest order, transport through the loop occurs through co-tunnelling of pairs of electrons with opposite spin, which is permitted due to the absence of a charge gap. Thus, when a flux is threaded through the loop, the zero-bias conductance will exhibit Aharonov-Bohm oscillations with period $h/2e$ characteristic of a superconductor rather than $h/e$ as would occur in a normal metal or Luttinger liquid. This behaviour is also reminiscent of the anomalous flux periodicity for Luther-Emery systems  which was demonstrated via finite-size bosonization in a previous study \cite{SeidelLee}. For identical reasons, if the density in the loop is tuned by gates and the loop sufficiently small to exhibit Coulomb blockade, charge sensing measurements will exhibit even-odd behaviour characteristic of superconductors, with the Coulomb energy being $\frac{E_C}{2} ( N - N_g)^2$  for even $N$ and $\frac{E_C}{2} ( N - N_g)^2 + \Delta$ for odd $N$.  Since the charging energy measured in existing nanowire systems is generally in the range $E_C \sim 10-100$ \textmu eV, observation of a spin gap $\sim 10$ \textmu eV is well within the capability of typical experiments. It should be noted that the gates must be placed symmetrically around the nanowire system in order to maintain the spatial symmetry of the screened two-electron interaction. The geometry also ensures the absence of any possible low-energy edge states which might permit single electron tunnelling.

\section{Conclusion}

The phenomenology of the class of models we have considered coincides with 1D systems with attractive interactions (e.g. the $U<0$ Hubbard model), and originates from electronic pairing despite the fact that the underlying interactions are repulsive. This behaviour arises instead from the spin texture in the cross-section of the wire which gives rise to phenomena otherwise expected in 1D systems with an attractive interaction. We may contrast the present results with previous studies of interactions in spin-orbit coupled systems in which inversion symmetry is lifted and such features are absent \cite{YuSC,MorozBarnesSC,ChengSC,SunSDWSOC}. This work shows that extremely similar physical setups in which the appropriate symmetries are restored will yield the realization of a peculiar strongly correlated phase with remarkable observable features.

\begin{acknowledgments} 
The author acknowledges M. Burrello, K. Flensberg, P. Krogstrup, H. Scammell, J. Ingham, and J. Paaske for important discussions. This work was supported by the Danish National Research Foundation and Microsoft Station Q.
\end{acknowledgments}

\bibliography{spingap}
\bibliographystyle{apsrev4-1}

\end{document}